\documentclass[preprint,12pt]{elsarticle}

\usepackage{blindtext}
\usepackage{xcolor}
\usepackage{epsfig}
\usepackage{amsmath,amssymb,bm}
\usepackage{enumerate}

\usepackage{lineno}
\usepackage{hyperref}
\journal{Physics Letters B}

\begin{document}

\begin{frontmatter}
\title{Extending the constraint for axion-like particles as resonances at the LHC and laser beam experiments
}

%Alternative: Extending searches for axion-like particles in light-by-light scattering

\author{C. Baldenegro$^a$, S. Hassani$^b$, C. Royon$^a$, L. Schoeffel$^{b,*}$}
\address{$a$ University  of Kansas, Lawrence, Kansas, U.S.}
\address{$b$ CEA Saclay, Irfu/DPhP, Gif-sur-Yvette, France}
\vspace{20mm}
\address{$^*$ c.baldenegro@cern.ch, samira.hassani@cern.ch, christophe.royon@cern.ch, laurent.olivier.schoeffel@cern.ch}

\begin{abstract}

\noindent
We study the discovery potential of axion-like particles (ALP), pseudo-scalars weakly coupled to Standard Model fields, at the Large Hadron Collider (LHC). Our focus is on ALPs coupled to the electromagnetic field, which would induce anomalous scattering of light-by-light. This can be directly probed in central exclusive production of photon pairs in ultra-peripheral collisions at the LHC in proton and heavy-ion collisions.
We consider non-standard collision modes of the LHC, such as argon-argon collisions at $\sqrt{s_{NN}} = 7$ TeV and proton-lead collisions at $\sqrt{s_{NN}} = 8.16$ TeV, to access regions in the parameter space complementary to the ones previously considered for lead-lead and proton-proton collisions. 
In addition, we show that, using laser beam interactions, we can constrain ALPs as resonant deviations in the refractive index induced by anomalous light-by-light scattering effects. If we combine the aforementioned approaches, ALPs can be probed in a wide range of masses from the eV scale up to the TeV scale.

\end{abstract}

%%\keywords{QED, Equivalent Photon Approximation, LHC}

\end{frontmatter}

%%%%%%%%%%%%%%%%%%%%%%%%%%%%%%%%%%
\section{Introduction}
%%%%%%%%%%%%%%%%%%%%%%%%%%%%%%%%%%

\noindent
Charge-parity (CP) violation is an important consequence of the
Standard Model (SM) of particle physics. Though CP violation is 
inherent in the construction of the SM,
there is a longstanding question of why quantum chromodynamics (QCD) seems to preserve CP symmetry, since in principle there could be CP violating terms in the QCD Lagrangian density.
Indeed, the strong CP problem is supported by the absence of a neutron electric dipole moment \cite{Afach:2015sja}. To solve it, scalar or pseudo-scalar complex fields, called axions, have been postulated~\cite{Peccei:1977ur,Peccei:1977hh,Peccei:1977np}. 
It has been speculated that cold axions could have been produced in abundance during the QCD phase transition in the early universe and that they may constitute one element of the cold dark matter~\cite{Arias:2012az}.
 Axions considered in these models have small masses (below the meV scale), and have been heavily constrained by dedicated axion helioscopes. 
More generally, pseudo-scalars coupled to SM particles, known as axion-like particles (ALP), appear in theories with spontaneously broken global, approximate, symmetries as pseudo Nambu-Goldstone bosons. For instance, ALPs appear in supersymmetric extensions of the SM or string theories~\cite{Witten:1984dg,Conlon:2006tq,Svrcek:2006yi,Arvanitaki:2009fg}. The focus of this letter is on ALPs with a coupling to photons. ALPs with these couplings have been heavily constrained for sub-eV masses, but for masses above the eV scale and up to the TeV scale, collider-based searches in electron-positron or hadron-hadron collisions are necessary. The problem of searching for these particles relying only on their coupling to photons aggravates in hadronic colliders, where the dominant interactions are nuclear in nature.

In this letter, we discuss the sensitivity to ALPs weakly coupled to the electromagnetic field in the context of light-by-light scattering in a wide domain in mass, benefitting from the  fact that the LHC can accelerate protons and heavy-ions \cite{us,dEnterria:2013zqi,Aaboud:2017bwk,Sirunyan:2018fhl}.
First, we expose how the LHC data can be used to extend the search for ALPs beyond what is currently accessible or expected for masses above a few GeV and up to a few TeV~\cite{Sirunyan:2018fhl,Knapen:2016moh,Baldenegro:2018hng}, with an emphasis on non-standard collision modes of the LHC. It has been shown that searches based on ultra-peripheral collisions in lead-lead collisions constrain ALPs between masses of 1 GeV to about 100 GeV~\cite{Knapen:2016moh}, whereas searches based on proton-proton collisions using the proton tagging technique can constrain ALPs masses between $\sim$ 500 GeV and 2 TeV~\cite{Baldenegro:2018hng}. These searches can be seen as the extreme ends of the high intensity frontier (lead-lead) and the high energy frontier (proton-proton) of photon-photon collisions at the LHC.

The goal is to find a way of constraining the parameter region untouched by the aforementioned searches by considering argon-argon and proton-lead collisions at the center-of-mass energy per nucleon pair of 7 TeV and 8.16 TeV respectively, which, due to the smaller minimum impact parameter between the hadrons participating in the ultra-peripheral collision, gives access to invariant masses of photon pairs as large as 400 GeV.

The letter is organized as follows. The general principles of photon physics at the LHC are presented in Sec.~\ref{LbL}, followed by the simulation framework and analysis strategy in Sec.~\ref{simulation} and Sec.~\ref{strategy} respectively, with our main results on the projections on the ALP--photon coupling and mass plane in Sec.~\ref{results}. Finally, in the last section ~\ref{laser} of this letter, we illustrate how in a very similar context but at low energies using laser beams, we can constrain ALPs of masses in the eV scale. This can be readily tested in existing facilities~\cite{Sarazin:2016zer,King:2012aw,Takahashi:2018uut,Bogorad:2019pbu}. Conclusions are drawn in Sec. \ref{conclusions}.

%%%%%%%%%%%%%%%%%%%%%  
\section{Scattering of light-by-light}\label{LbL}
%%%%%%%%%%%%%%%%%%%%%

\begin{figure}[!htbp]
\centering
  \includegraphics[scale=0.4]{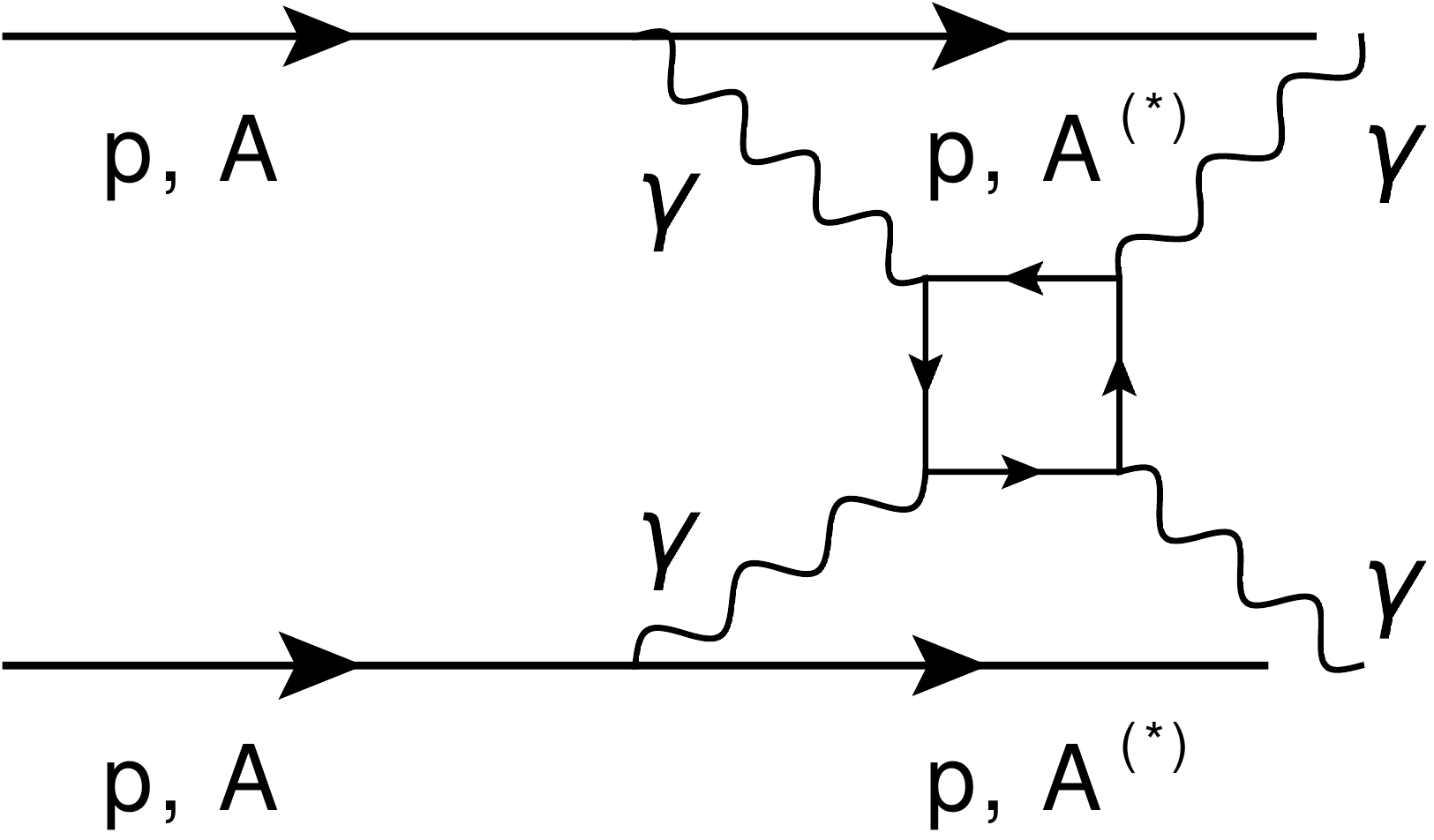}
  \caption{Schematic diagram of light-by-light scattering in the SM in proton-proton, proton-ion or ion-ion collisions. The box diagram includes quarks, charged leptons and $W$ boson contributions.}
  \label{diag1}
\end{figure}

\noindent
The scattering of light-by-light $\gamma\gamma \rightarrow \gamma\gamma$
%\begin{equation}
%\gamma\gamma \rightarrow \gamma\gamma
%\label{toto1}
%\end{equation}
can be probed in proton-proton, proton-ion or ion-ion collisions, where the initial state photons %(left hand side of reaction (\ref{toto1}))
 correspond to electromagnetic fields (EM) produced by the ultra-relativistic incident hadrons (protons or ions). 
These interactions occur in ultra-peripheral collisions (UPC), where the two incident hadrons do not collide centrally but are separated in the transverse direction by at least the sum of their radii. In the SM, the scattering of light-by-light is induced at one-loop at leading order (quark, charged leptons and $W$ boson contributions), as illustrated in Fig. \ref{diag1}.

The signature of these reactions is the presence of two photons and no additional activity in the central detector. The outgoing hadrons (protons or ions) escape into the beam pipe. In the case of protons, if they remain intact after the interaction, they can be tagged with dedicated forward proton spectrometers installed at the LHC. The outgoing heavy-ions may be affected by giant dipole resonances, followed by neutron emission and excited states of the heavy-ion~\cite{Auerbach:1983hld}.
Lead-lead collisions are an ideal probe of elastic scattering of photons, since the effective photon luminosity scales with $Z^4$, where $Z$ is the electric charge of the colliding ions.
 The cross section is then considerably enhanced in $PbPb$ collisions (with $Z=82$ protons and $A=208$ nucleons for the lead ions used at the LHC) by a factor $82^4$ compared to $pp$ collisions. Indeed, evidence for the scattering of light-by-light was presented by the ATLAS and CMS collaborations in lead-lead collisions ~\cite{Aaboud:2017bwk, Sirunyan:2018fhl}, paving the path for future analyses in heavy-ion collisions.

\subsection{Axion-like particles in light-by-light scattering}
ALPs of masses above a few GeV and weakly coupled to other particles of the SM can be probed at the LHC. Their presence would be seen as a deviation on the elastic scattering of light-by-light ~\cite{Sirunyan:2018fhl,Knapen:2016moh,Baldenegro:2018hng}. In this context, ALPs could be produced as intermediate states in the reaction: 
\begin{equation}
\gamma \gamma \rightarrow a \rightarrow \gamma \gamma,
\label{toto2}
\end{equation}
thus modifying the di-photon invariant mass distribution. Indeed, an ALP of mass $m_a$ would appear as a resonance in the distribution of the invariant mass of the photon pair. 
Therefore, the search for ALPs using reaction (\ref{toto2}) is based of the possibility to find a deviation in the di-photons mass spectrum modulo the uncertainty of the light-by-light prediction.

\begin{figure}[!htbp]
\centering
  \includegraphics[scale=0.4]{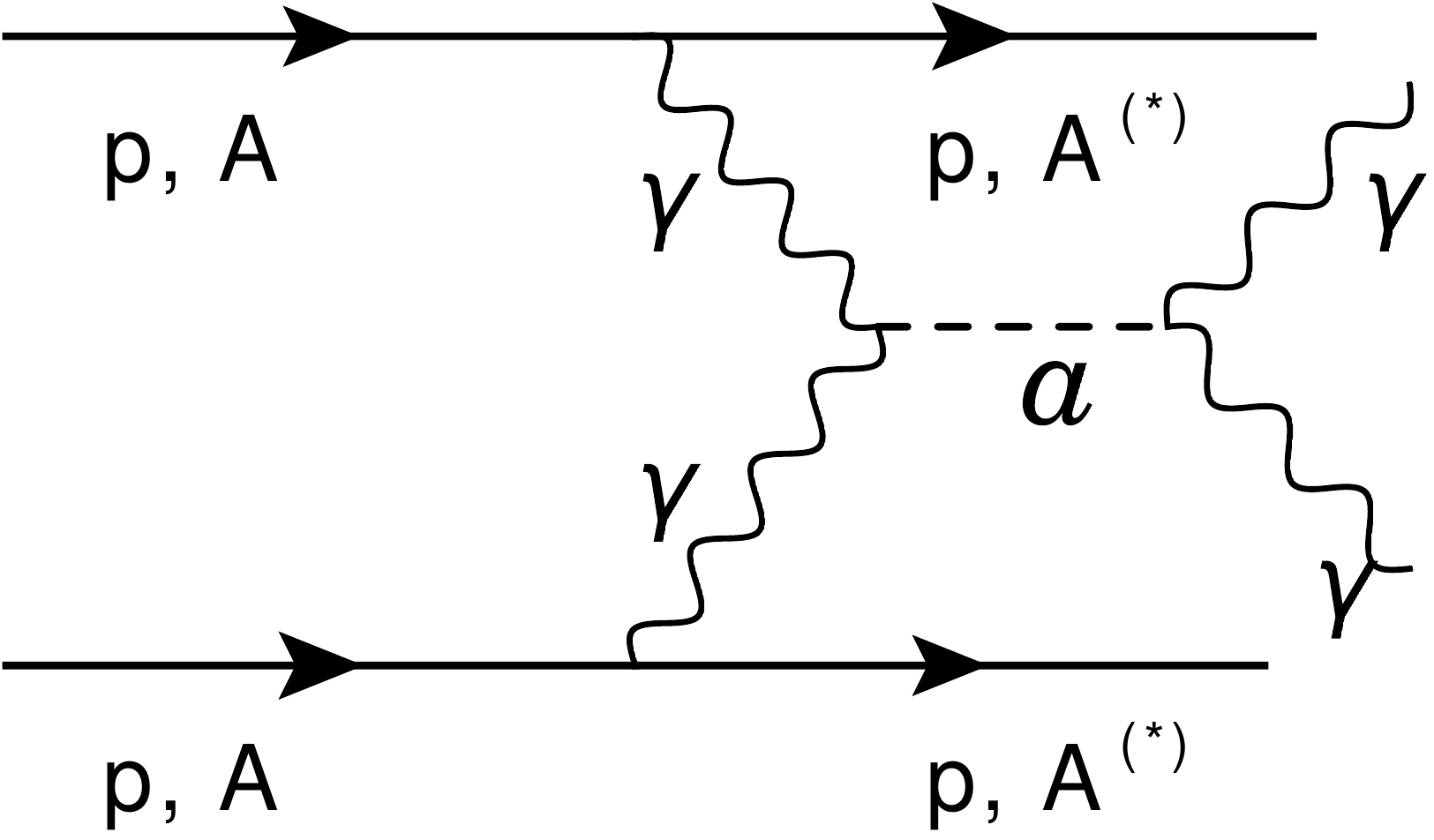}
  \caption{Schematic diagram for central exclusive production of an axion-like particle in photon-induced processes in proton-proton, proton-ion or ion-ion collisions.}
  \label{diag2}
\end{figure}

%%%%%%%%%%%%
\subsection{Experimental context at the LHC}
%%%%%%%%%%%%

\noindent
Experimentally, at the LHC, the first limits have been obtained using data collected in lead-lead ($PbPb$) collisions at the LHC with a center-of-mass energy per nucleon pair of $5.02$ TeV \cite{Sirunyan:2018fhl,Knapen:2016moh} from masses of ALPs of $5$~GeV up to $100$ GeV.
One important limitation in $PbPb$ collisions is that there exists maximum energy for the incoming photons (emitted by the relativistic incident $Pb$) which is of the form $\gamma/R_{Pb}$, where $\gamma$ is the Lorentz factor and $R_{Pb}$ the transverse size of the incident ions. Indeed, the need for a minimum distance separation between the two hadrons for a UPC to take place sets an upper bound on the invariant mass of the final two photons, which is inversely proportional to the size of the incident particles. Thus, larger di-photon masses could be probed
with ions of smaller sizes, or even better in $pp$ collisions
\cite{Baldenegro:2018hng}. Experimentally, such measurements in $pp$ collisions are achievable if protons are tagged in dedicated forward proton spectrometers (very far from the interaction point of the reaction), otherwise the detection of the rapidity gap is problematic due to the large number of interactions per bunch crossing. The production rate is much smaller than in the heavy ion collisions case, but the access to higher invariant masses, and the growth of the cross section for exclusive ALP production with the invariant 
mass of the di-photon system, leads to sensitivities on the ALP-photon coupling comparable in magnitude to the ones achievable in $PbPb$ collisions. For the moment, this is only a prospective but the interest is that masses of ALPs above the TeV range could be further tested this way ~\cite{Baldenegro:2018hng}.

%%%%%%%%%%%%
\subsection{Prospects at the Large Hadron Collider}
%%%%%%%%%%%%

\noindent
According to these first results and prospects using light-by-light scattering, it seems very logical to consider what could be obtained with larger periods of data taking in $PbPb$ collisions, thus increasing the statistics already recorded
and accessing larger masses for ALPs searches. In the same spirit and following the arguments presented above, it is possible
to probe intermediate mass ranges above $50$ GeV and below $1$ TeV by using 
proton-lead collisions or different species of ions of smaller sizes than lead nuclei, like argon ($Ar$) ($A=40$, $Z=18$)~\cite{Citron:2018lsq}. Argon-argon collisions have attracted the attention of the nuclear physics community to obtain, for instance, additional constraints on nuclear parton distribution functions in lighter nuclei and study collective phenomena in smaller systems. Beams of argon-argon allow the possibility of collecting larger luminosities than in the lead-lead or proton-lead collision modes, as can be seen in detail in Ref.~\cite{Citron:2018lsq}. It is useful to see what other uses can be made of the argon-argon collisions, in addition to the ones exposed in the context of the understanding of nuclear matter in Ref.~\cite{Citron:2018lsq}.
The purpose of this letter is to derive new projections in this broader context, studying exclusive production of two photons 
in ion-ion, proton-ion and proton-proton UPC collisions.

%%%%%%%%%%%%%%%%%%%%
\section{Simulations of physics processes at the LHC}\label{simulation}
%%%%%%%%%%%%%%%%%%%%

\noindent
The light-by-light scattering and ALPs production (as exposed in the previous section) in ion-ion, proton-ion and proton-proton collisions at the LHC have been simulated using the Forward Physics Monte Carlo (FPMC) event generator  \cite{Boonekamp:2011ky}, which is a dedicated generator for photon-induced and diffractive processes. Originally, the FPMC generator was introduced to handle reactions involving protons or anti-protons collisions, but we have introduced the possibility of handling heavy-ion collisions as well, as described below.
We provide the description explicitly for the ion-ion case, while the same logic applies to the proton-ion case, which is also encoded in the new version of FPMC~\footnote{The most updated version of the FPMC event generator is available in \href{github.com/fpmc-hep/fpmc}{github.com/fpmc-hep/fpmc}, with a release note in preparation by the authors.}.
The description of of the photon-induced reactions in ion-ion collisions is based on the equivalent photon approximation applied to ultra-relativistic ions and 
where the equivalent photons are produced at low virtualities.
This follows what is done in other event generators \cite{Klein:2016yzr,Klusek-Gawenda:2016euz}.
The cross section for the collision of two ions $A$, with the same number of nucleons, reads:
\begin{equation}
\sigma(A+A \rightarrow A+A+\gamma +\gamma) = 
\int
n(\vec{b}_{1},\omega_1) n(\vec{b}_{2},\omega_2) 
\sigma_{\gamma \gamma \rightarrow \gamma \gamma}(\omega_1,\omega_2) 
 \frac{d\omega_1}{\omega_1}  \frac{d\omega_2}{\omega_2}
 d^2 \vec{b}_{1} d^2 \vec{b}_{2}
\label{lbl1}
\end{equation}
where $\vec{b}_1$, $\vec{b}_2$ are the impact parameters (ion centers to the interaction point), $\omega_{1,2}$ the energies of the incoming photons,
$n(\vec{b}_{1,2},\omega_{1,2})$ the photon-flux functions and $\sigma_{\gamma \gamma \rightarrow \gamma \gamma}$ the cross section for the light-by-light scattering itself, computed as in Ref. \cite{Baldenegro:2018hng}.
The integration has some constraints: 
(1) in order to impose that the reaction is ultra-peripheral (and not inelastic): $b_1, b_2>R_A$ and $|\vec{b}_1-\vec{b}_2|> 2R_A$ (non overlap), where $R_A$ is the ion radius,
(2) the reaction can happen only with some requirements on the photons momenta: typically, the virtuality of each photon $Q^2$
 needs to be smaller than $1/R_A^2$ in order to probe it as a whole electrically charged object. Then, we can consider the action of all the charges in the nucleus to be coherent and a factor $Z$ will appear in the EM fields, leading to a $Z^4$ factor in the cross section (see above). 
 Let us comment the requirement of non-overlap, $|\vec{b}_1-\vec{b}_2|> 2R_A$. In general, if we call $\Omega(b)$ the opacity function of the heavy-ion, then the above condition is equivalent to:
$$
e^{-\Omega( |\vec{b}_1-\vec{b}_2|)} = \Theta(|\vec{b}_1-\vec{b}_2|-2R_A),
$$
where $e^{-\Omega(b)}$ is the probability of non-inelastic interaction in the ion-ion collisions and $\Theta$ is the Heaviside step function. In principle, it is possible to define this function more accurately with a form which turns smoothly from $0$ to $1$ at $b=2R_A$. However, in the case of UPCs considered here, the approximation as a Heaviside step function is very good. There is only one case, the QCD exclusive production of two photons induced by two-gluons exchange for which the impact parameter of the reactions may be small: for this case, we use a dedicated Monte Carlo generator. \

Coming back to equation (\ref{lbl1}),
 we can write:
$$
n(\vec{b}_{1},\omega_1) n(\vec{b}_{2},\omega_2) =
\frac{1}{\pi} \left | \vec{E}_\mathrm{T} (\omega_1,\vec{b}_1)  \right |^2
\frac{1}{\pi} \left | \vec{E}_\mathrm{T} (\omega_2,\vec{b}_2)  \right |^2
$$
where the transverse EM fields are the fields produced by the ultra-relativistic ions $A, Z$:
$$
\vec{E}_\mathrm{T} (\vec{b},\omega) = Z|e|
\int \frac{d^2 \vec{k}_\mathrm{T}}{(2 \pi)^2} \ e^{-i \vec{b} \vec{k}_\mathrm{T}}
 \frac{(-i\vec{k}_\mathrm{T})}{k_\mathrm{T}^2+\frac{\omega^2}{\gamma^2}}
 F(k_\mathrm{T}^2+\frac{\omega^2}{\gamma^2}).
$$
In this expression, there are several dependencies on the ion type
($A, Z$):
the trivial dependence in $Z$, a second dependence in the integration limit $k_\mathrm{T} < 1/R_A$ and another dependence in the electromagnetic form factor $F(.)$ for the ion, which includes the electric and magnetic components. 
Let us note that for ion-ion collisions, the magnetic component leads to a negligible contribution to the cross section (\ref{lbl1}).
In FPMC, we have encoded the monopole expression and realistic form factors (Fourier transform of the charge distribution of the ions)  \cite{Klusek-Gawenda:2016euz}. In the following, we use the realistic form factors. Then, the integration in equation (\ref{lbl1}) can be done easily. 

In the following, we use the FPMC generator for the generation of all photon-induced processes. Processes originating from two-gluon exchanges are generated with the dedicated SuperChic v3 Monte Carlo event generator ~\cite{Harland-Lang:2018iur}, as explained in the next section. Based on these modifications of the code, the cross section for the light-by-light scattering cross sections in $PbPb$ at a center-of-mass energy of $5.02$ TeV per nucleon pair is found to be $230$ nb for a minimum transverse momentum $p_\mathrm{T}$ of each outgoing photon of $2$ GeV. 

For the computation of cross sections for ALPs production
$\sigma_a$, the same procedure is used with:
\begin{equation}
\sigma_a = 
\int
n(\vec{b}_{1},\omega_1) n(\vec{b}_{2},\omega_2) 
\sigma_{\gamma \gamma \rightarrow a \rightarrow \gamma \gamma}(\omega_1,\omega_2) 
 \frac{d\omega_1}{\omega_1}  \frac{d\omega_2}{\omega_2}
 d^2 \vec{b}_{1} d^2 \vec{b}_{2},
\label{alp1}
\end{equation}
where the evaluation of $\sigma_{\gamma \gamma \rightarrow a \rightarrow \gamma \gamma}$ follows the method defined in Ref. \cite{Baldenegro:2018hng}. 
For our studies, we model the coupling of the pseudo-scalar field $a$ of mass $m_a$ with the EM fields via the interacting Lagrangian density:
$$
L_{int} = \frac{1}{f} a F_{\mu \nu}\tilde{F}_{\mu \nu},
$$
where $1/f$ is the ALP-photon coupling, $F_{\mu \nu}$ is the EM field tensor and $\tilde{F}_{\mu \nu}$ is the dual tensor of the EM field tensor. The ALP production cross section is then computed in the narrow resonance approximation with a minimal decay width:
$$
\Gamma(a \rightarrow \gamma \gamma) = \frac{m_a^3}{4\pi f^2}.
$$

\begin{table}
\begin{center}
\begin{tabular}{|c|c|c|}
\hline
   $m_a$ in GeV & $\sigma_a(PbPb)$  & $\sigma_a(PbPb)$ \\
    & STARlight& FPMC \\
   \hline
   30 & $533$ nb  &  $536$ nb  \\
   50 & $170$ nb  & $174$ nb  \\
   100 & $18$ nb  & $18.5$ nb  \\
   200 & $481$ pb &  $485$ pb  \\
   \hline
\end{tabular}
\end{center}
\caption{Cross sections for various masses of ALPs production in $PbPb$ collisions for a center-of-mass energy per nucleon pair $5.02$ TeV. We have taken a coupling value of $f=4$ TeV for explicit comparison.}
\label{t2}
\end{table}

As in Ref. \cite{Baldenegro:2018hng}, in the forthcoming projections, the decay width of $a$ is taken as a free parameter satisfying $\Gamma>\Gamma(a \rightarrow \gamma \gamma)$ and the branching ratio is defined as
$\mathcal{B}(a \rightarrow \gamma \gamma)=\Gamma(a \rightarrow \gamma \gamma)/\Gamma$. In this context, the ALP production cross section is proportional to $\frac{1}{f^2} \mathcal{B}(a \rightarrow \gamma \gamma)$.

We have checked that the results obtained for 
the generation of the ALPs processes is compatible with the  STARlight MC generator
\cite{Klein:2016yzr} (see Tab. \ref{t2}).

%%%%%%%%%%%%%%%%%%%%
\section{Analysis strategy and detector effects at the LHC}\label{strategy}
%%%%%%%%%%%%%%%%%%%%

\noindent
The analysis strategies for ion-ion and proton-ion 
 follow  the principles and strategies described in Ref. \cite{Aaboud:2017bwk, Sirunyan:2018fhl, Knapen:2016moh,Baldenegro:2018hng}. 
Our selection focuses mainly on the two photons detected by the central detectors, which are reconstructed as two isolated EM clusters and identified as photons.
 We emulate the detector resolutions for the energies and angles of the outgoing photons, as well as the $p_T$ dependent reconstruction efficiencies according to Ref. \cite{Aad:2009wy}. The overall reconstruction efficiency is about $80$ \% for multi-GeV photons and the resolution ranges from $5$ \% for low masses of photons (below $15$ GeV) down to $1$~\% for larger masses (above $50$ GeV).
 
 The ion-ion and proton-ion collisions considered here are the following:
 \begin{itemize}
\item $PbPb$ collisions at a center-of-mass energy per nucleon pair of $5.02$~TeV, with a luminosity of $10$~nb$^{-1}$,
\item $ArAr$ collisions at a center-of-mass energy per nucleon pair of $7$~TeV, with a luminosity of $3$~pb$^{-1}$ and
\item $pPb$ collisions at a center-of-mass energy per nucleon pair of $8.16$~TeV, with a luminosity of $5$~pb$^{-1}$.
\end{itemize}

See Ref. \cite{Citron:2018lsq} for a justification of these choices,
which correspond to the future opportunities of the heavy-ion program at the LHC.
For all physics cases, 
the analysis is based on the following requirements: we request two photons in the final state, within pseudo-rapidity values of $|\eta|<2.4$, minimum transverse momentum of $p_\mathrm{T} > 3$ GeV for each photon, an invariant mass above $6$ GeV, and the magnitude of the di-photon transverse momentum to be at most  $p_\mathrm{T}^{\gamma\gamma}<2$ GeV and an acoplanarity ($|\Delta \phi^{\gamma\gamma}/\pi-1|$) below $0.01$ units. The last two requirements select mainly central exclusive production processes, which tend to be strongly correlated in momentum due to the absence of underlying event activity.
{%\color{red}
Also, the invariant mass resolution for the two outgoing photons varies from $0.5$ GeV at low masses (below $15$ GeV) up to $1$ GeV for larger
masses.}

\begin{figure}[!htbp]
\centering
  \includegraphics[scale=0.5]{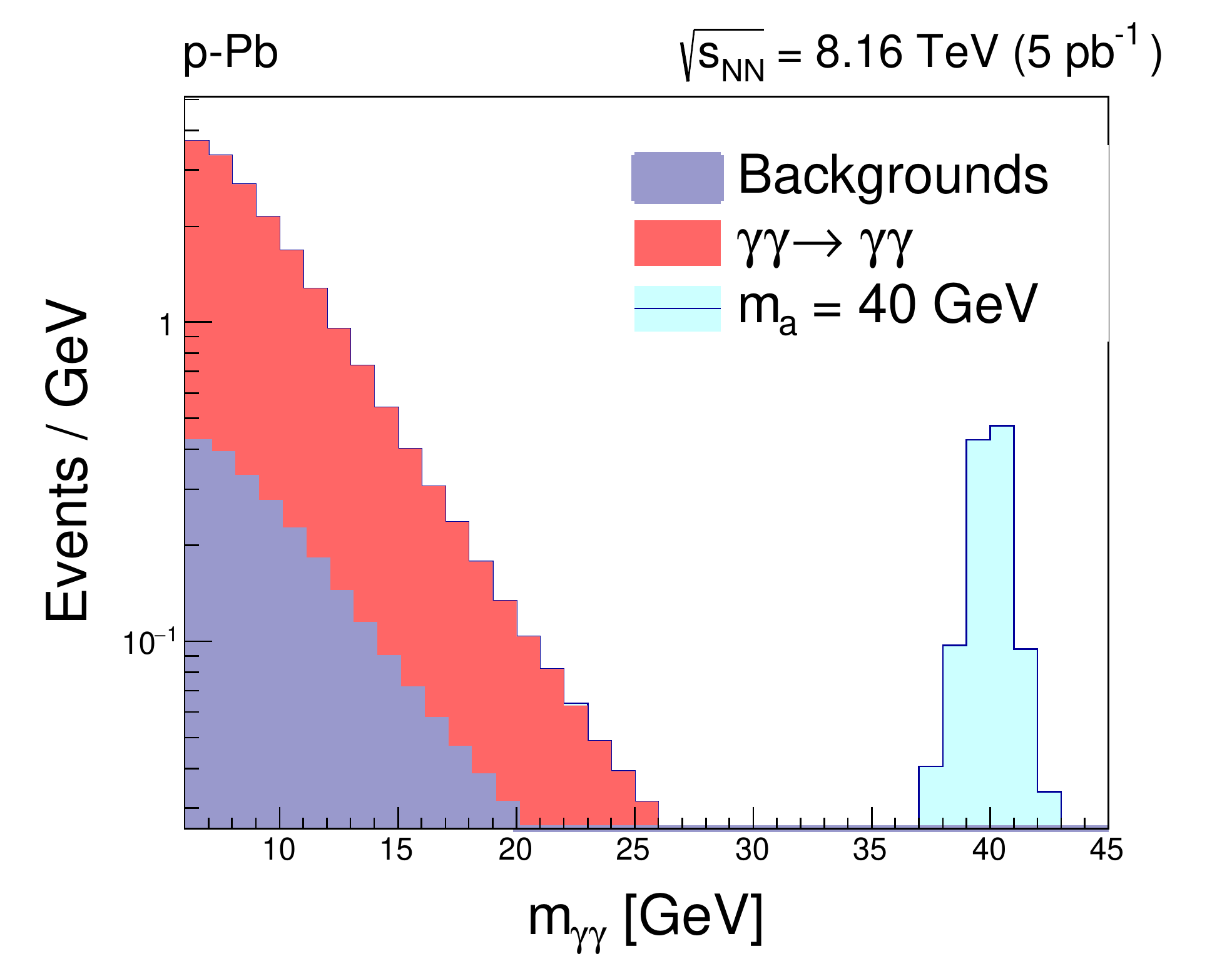}
  \caption{Di-photon invariant mass
distribution in ultra-peripheral $pPb$ collisions at $\sqrt{s}_{NN} = 8.16$ TeV. We show explicitly the SM light-by-light scattering contribution (red) and other subleading background processes (lavender). For illustrative purposes, we show an instance of an ALP signal with mass $m_a = 40$ GeV and a coupling $f = 2$ TeV. See text for more details.}
  \label{fig1}
\end{figure}

There are two types of backgrounds important for ALP searches as exclusive photon pair production. The dominant irreducible background comes from the SM light-by-light scattering process (described in the previous sections). 
The other irreducible background arises from the $two$-gluons exchange process between the colliding ions or the ion and the proton, called central exclusive process (CEP) of photons.
The contribution of this process is evaluated using the SuperChic v3 Monte Carlo generator ~\cite{Harland-Lang:2018iur}.
The contribution of CEP has been found to be negligible for invariant masses of the two outgoing photons above $15$ GeV, while it is found to be below $4$ \% of the light-by-light scattering contribution for lower masses, when the selection requirements described above are applied.
Finally, there is the central exclusive production of electron-positron production, where the pair can be mis-identified as a photon pair.  This contribution has been found to be negligible for invariant masses of the two outgoing photons above $15$ GeV and below $5$ \% for lower masses,
once the selection requirements described above are applied. 
For illustration, in Fig. \ref{fig1} we present the invariant mass
distribution of the two outgoing photons in the case of $pPb$ collisions at $8.16$ TeV, for an ALP of mass equal to $40$ GeV and coupling $f = 2$ TeV,
together with contributions of light-by-light and other background processes (see above).

%%%%%%%%%%%%%%%%%%%%%
\section{Results}\label{results}
%%%%%%%%%%%%%%%%%%%%%

\begin{figure}[!]
\centering
\hspace*{-2.3cm}
\includegraphics[scale=0.45]{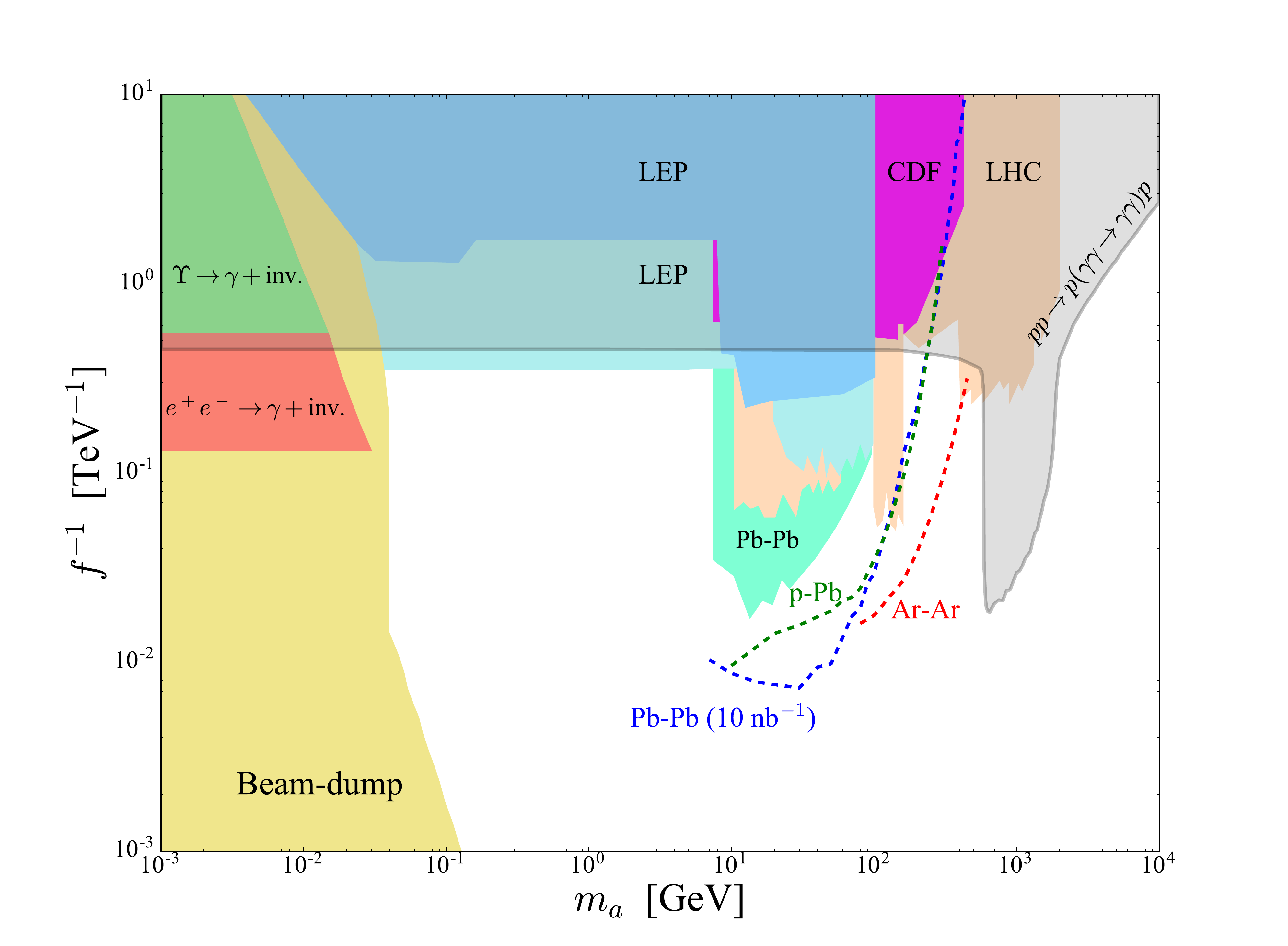}
  \caption{Exclusion limits on the ALP-photon coupling as a function of the ALP mass derived from several particle physics and astro-particle physics experiments (see text and Ref. \cite{Bauer:2017ris,Jaeckel:2015jla}).
  Projections for $pp$ collisions with the proton tagging technique are drawn from Ref. \cite{Baldenegro:2018hng}.
   The projections derived in this letter for
$pPb$ collisions (at $8.16$ TeV per nucleon pair and for a luminosity of $5$~pb$^{-1}$),
$PbPb$ collisions 
(at $5.02$ TeV per nucleon pair and for a luminosity of $10$~nb$^{-1}$) 
(in agreement with Ref. \cite{Knapen:2016moh})
and
 $ArAr$ collisions 
(at $7$ TeV per nucleon pair and for a luminosity of $3$~pb$^{-1}$)  are  shown as dotted lines, under the assumption that
   $\mathcal{B}(a \rightarrow\gamma\gamma)=1$. }
  \label{superTOTO}
\end{figure}

\noindent
The existing bounds presented in Fig. \ref{superTOTO} were originally compiled in Ref. \cite{Bauer:2017ris, Jaeckel:2015jla}.
Fig. \ref{superTOTO} illustrates a subset of these bounds for ALPs masses between 10$^{-3}$ GeV and 2 TeV and coupling values to photons as low as $10^{-3}$ TeV$^{-1}$, although some bounds exist down to values of 10$^{-15}$ GeV in ALP mass and 10$^{-10}$~TeV$^{-1}$ in the ALP-photon coupling. On this figure, we note that ALPs have been constrained in beam dump searches, which probe resonant production of neutral pseudoscalar mesons in photon interactions with nuclei, known as the Primakoff effect. Various beam-dump experiments at SLAC collectively yield the area in yellow \cite{PhysRevLett.59.755,PhysRevD.38.3375,Döbrich2016}. Also, Upsilon meson decays searched at the CLEO and BaBar experiments \cite{PhysRevD.51.2053,delAmoSanchez:2010ac} exclude the region shaded in green.
Collider-based constraints are derived by recasting results on mono-photon, di-photon and tri-photon searches  at LEP (light blue corresponds to Run 2 of LEP whereas dark blue corresponds to Run 1 of LEP), the Tevatron (magenta, by CDF) and the LHC (peach for $pp$ collisions). 
Finally, Fig. \ref{superTOTO} illustrates 
 the more recent results on light-by-light scattering by the ATLAS and CMS collaborations in $PbPb$ collisions (green)~\cite{Aaboud:2017bwk, Sirunyan:2018fhl}. Let us note that collider-based bounds assume ~$\mathcal{B}(a \rightarrow\gamma\gamma)=1$.

Then, we present new projections for ion-ion and proton-ion collisions
derived following the analyses exposed in the previous section.
The di-photon invariant mass distribution is the key discriminating variable since the ALP would manifest as a resonant bump over the smooth spectrum,
with bin widths comparable to the expected resolution of a narrow resonant ALP signal. 
The light-by-light scattering in the SM, misidentified exclusive electron-positron pair production and CEP of photon pairs are considered as backgrounds in
this search.
A binned likelihood
function is constructed in each bin of the invariant mass distribution from the Poisson probability of the sum of the
contributions of the background and a hypothetical signal of strength relative to the benchmark model.
This likelihood function is used to set limits on the presence of a signal. A systematic uncertainty of
$30$ \% is considered for the shape of the  background distribution as well as a systematic uncertainty
of $20$ \% on its normalization. 
The systematic uncertainties enter as nuisance parameters with Gaussian
or log-normal prior distributions, in convolution with the nominal background distribution.
Upper limits are set on the product of the production cross section of new ALPs resonances and their decay
branching ratio into two photons. Exclusion intervals are derived using the CLs method in the asymptotic
approximation \cite{Read:2002hq}. The limit set on the signal strength is translated into a limit on the signal cross section times branching ratio and the coupling.

%%%%%%%%%%%%
In Fig. \ref{superTOTO}, our projections extracted for $pPb$ collisions (at $8.16$~TeV per nucleon pair and for a luminosity of $5$~pb$^{-1}$),
$PbPb$  collisions (at $5.02$~TeV per nucleon pair and for a luminosity of  $10$~nb$^{-1}$)
 and
$ArAr$  collisions (at $7$~TeV per nucleon pair and for a luminosity of  $3$~pb$^{-1}$)
are superimposed as dotted lines to the previous observed limits discussed above.
Argon-argon collisions have the potential of providing stringent constraints on the ALP-photon coupling for masses between 100 GeV to about 400 GeV, covering a region of parameter space that is rather difficult to access in proton-proton or lead-lead collisions. Projections derived from proton-lead collisions have the potential of covering larger invariant masses relative to lead-lead collisions, but unfortunately are heavily limited by the expected amount of luminosity in this mode. Argon-argon collisions have a better possibility of providing competitive bounds on the ALP-photon coupling in a kinematic region rather difficult to cover by other probes. 
{%\color{red} 
Let us note that
the future High Luminosity LHC, with a planned integrated luminosity 3 ab$^{-1}$ in $14$ TeV $pp$ collisions, has great potential of providing stringent constraints on ALPs coupled to photons over a wide range of masses, especially with the upcoming Phase-2 upgrades of the LHC experiments. This can be done, for instance, in inclusive tri-photon production ($q\bar{q}\rightarrow a\gamma$) \cite{Mimasu:2014nea}. Our projections in $ArAr$ collisions yield similar sensitivities to the ones in inclusive tri-photon production in $pp$ collisions for ALP masses at about $100$--$200$ GeV, and can be seen as an alternative probe of ALPs in this regime.

}

%%%%%%%%%%%%

%%%%%%%%%%%%%
\section{Ideas for ALPs of small masses in the low energy domain}\label{laser}
%%%%%%%%%%%%%

%%%%%%%%%%%%
\subsection{Position of the problem}
%%%%%%%%%%%%

\noindent
As mentioned in the introduction, the link between ALPs and light-by-light scattering is not restricted to the high energy domain. We develop some ideas in this sense below. Let us note that despite
the experimental observation of light-by-light scattering at the LHC
 was an important success, a direct observation using real photons still remains to be reported, using laser beams to generate the EM fields \cite{Sarazin:2016zer,King:2012aw,Takahashi:2018uut,Bogorad:2019pbu}. 
In this context, even with high intensity lasers of $10^{24}$ W/cm$^2$, these EM fields would be much smaller than their critical values ($E\ll10^{18}$ V/m). Then, the non-linearity of the theory that produces the scattering of light-by-light can be approximated by the effective Euler-Heisenberg Lagrangian density:
\begin{equation}
{\mathcal L} 
= \frac{1}{2} (\textbf{E}^2-\textbf{B}^2)+ 
\frac{2}{45}\frac{\alpha^2}{m_e^4} \left[
(\textbf{E}^2-\textbf{B}^2)^2 +7(\textbf{E}.\textbf{B})^2
\right],
\label{eh}
\end{equation}
where $\alpha$ is the fine structure constant of EM interactions and $m_e$ the mass of the electron.
The second term in this expression, bi-quadratic in the EM fields, gives the size of the non-linearities. The light-by-light cross section in this approximation is thus proportional to 
$[\frac{\alpha^2}{m_e^4}]^2 \omega^6$, where $\omega$ is the angular frequency, taken to be the same for both beams in this expression. In practice, in the recent years, there has been a large interest in testing this non-linearity in vacuum by measuring the scattering of intense laser beams
\cite{Sarazin:2016zer,King:2012aw,Takahashi:2018uut,Bogorad:2019pbu}. In a specific experimental configuration, we show  that, at the interaction area of the two laser beams, the non-linear coupling of the EM fields produces an increase of the vacuum refractive index, which then produces the scattering of the probe beam. The contribution of ALPs would modify this effect in a resonant way.

%%%%%%%%%%%%
\subsection{Two counter-propagating waves and the search of ALPs}
%%%%%%%%%%%%

\noindent
First, we study the interaction of two
counter-propagating EM waves along the $(z)$ axis produced by high intensity laser beams. These two laser beams are produced in high finesse cavities.
For simplicity, we develop the calculations for plane waves of different intensities (linearly polarized), defined as:
$$ 
\textbf{E}_0 = E_0 \textbf{e}_x e^{i\omega_0 (t+z)} +c.c.
$$
and 
$$ 
\textbf{E}_1 = (E_{1,x} \textbf{e}_x+E_{1,y} \textbf{e}_y) e^{i\omega (t-z)} +c.c.
$$ 
with $E_0\gg E_{1,x},E_{1,y}$
and different angular frequencies $\omega_0 \ne \omega$. 
Without loss of generality, we can pose: $E_0= \mathrm{constant}$ and $E_{1,x}=E_{1,x}(z)$, $E_{1,y}=E_{1,y}(z)$, where the $z$ dependence for $E_{1,x}$ and $E_{1,y}$ will be a consequence of the bi-quadratic  terms of the EM fields in expression (\ref{eh}).  
Introducing a transverse profile to the EM fields would not modify the conclusions exposed below.
We obtain $E_{1,x}(z)$ and $E_{1,y}(z)$ by solving the non-linear Maxwell equations derived from the Lagrangian density (\ref{eh}), namely:
\begin{equation}
\nabla \times \textbf{E}  = -\partial_t{\textbf{B}},
\label{eqb}
\end{equation}
\begin{equation}
\nabla \times \textbf{H} = \partial_t{\textbf{D}},
\label{eqa}
\end{equation}
with the vectors $\textbf{D}$ and $\textbf{H}$ defined as:
\begin{equation}
\textbf{D}=\textbf{E}+ K\left[2\left(\textbf{E}^2-\textbf{B}^2\right)\textbf{E}+7\left(\textbf{E}\cdot\textbf{B}\right)\textbf{B}\right],
\label{eqD}
\end{equation}
\begin{equation}
\textbf{H}=\textbf{B}+ K\left[2\left(\textbf{E}^2-\textbf{B}^2\right)\textbf{B}-7\left(\textbf{E}\cdot\textbf{B}\right)\textbf{E}\right],
\label{eqH}
\end{equation}
where $K=\frac{16\pi}{45}\frac{\alpha^2}{m_e^4}$. First, the magnetic field 
$\textbf{B}$ is obtained from equation (\ref{eqb}). Then,
$\textbf{E}$ and $\textbf{B}$ are used in equations (\ref{eqD}) and (\ref{eqH}) in order to derive the vector fields $\textbf{D}$ and $\textbf{H}$. Finally, by solving equation (\ref{eqa}), we obtain:

\begin{equation}
E_{1,x}(z) = E_{1,x}(0) e^{i 16 K \omega |E_0|^2 z},
\ \ \
E_{1,y}(z) = E_{1,y}(0) e^{i 28 K \omega |E_0|^2 z}.
\end{equation}

\noindent
The interaction of the two counter-propagating EM wave through the creation of virtual electron-positron pair, leading to light-by-light scattering, generates a birefringence of the vacuum. 
Then, the non-linear coupling of the fields induces an increase of the vacuum refractive index.
In the physics case considered here, the refractive index along the strong electric field is found to be: $n_x = 1+16 K  |E_0|^2$, while the index in the perpendicular direction is: $n_y = 1+28 K  |E_0|^2$.

Experimentally, let us consider that the wave $\textbf{E}_1$ is
initially polarized with an angle $\alpha$ w.r.t. the $(x)$ axis
and then interact with the counter propagating intense wave $\textbf{E}_0$ over a length $L$. After the propagation over $L$:
$$
\textbf{E}_1(L,t) = (A \textbf{e}_x \cos \alpha e^{in_x \omega L}
+A \textbf{e}_y \sin \alpha e^{in_y \omega L}) e^{i\omega t} + c.c.
$$
There will be a field component perpendicular to the initial polarization and the intensity of this mode is proportional to $\sin^2 2\alpha \sin^2\theta$
with 
$$
\theta = \pi (n_y-n_x)L/\lambda.
$$
Numerically, for a field of high intensity of about $10^{24}$ W/cm$^2$ focused on a cross section area of $10^{-6}$ cm$^2$
with $\lambda=0.5$ $\mu$m, this gives: $\theta \sim 10^{-8}$ rad.

At this step, 
we can study how this result is modified 
in the presence of ALPs, considered as a pseudo-scalar field $a(x,y,z,t)$ 
of mass $m_a$  with a coupling to the EM fields of the form:
${\mathcal L}_{int}=- {\eta} \ a \textbf{E} \cdot \textbf{B}$,
where $\eta \equiv 4/f$ following the notations of the previous sections.
The coupling of the fields $a$, $\textbf{E}$ and $\textbf{B}$ is then encoded in two equations:
\begin{equation}
(\square +m_a^2)a = -\eta  \ \textbf{E} \cdot \textbf{B},
\label{phi1}
\end{equation}
\begin{equation}
\nabla \times \textbf{H} = \partial_t{\textbf{D}} + {\eta} \ [\textbf{E} \times 
\nabla a - \textbf{B} \frac{\partial a}{\partial t}].
\label{phi2}
\end{equation}
Equation (\ref{phi1}) is the equation of motion of field $a$ coupled to the EM fields, while equation (\ref{phi2}) is the modified form of equation (\ref{eqa}), when ALPs are taken into account.

In equation (\ref{phi1}), we see that the right-hand side of the equation contains terms in $e^{\pm i(\omega_0+\omega)t}$ and terms in $e^{\pm i(\omega_0-\omega)t}$, which correspond to possible time dependence for the ALP field. We need to keep both terms in the calculations \cite{scott}.
Then, the equation of motion for $a$ (\ref{phi1}) gives:
\begin{equation}
(\square +m_a^2)a = 2\eta  e^{-i(\omega_0-\omega)t-i(\omega_0+\omega)z} \bar{E_0}E_{1,y}+
2\eta  e^{i(\omega_0+\omega)t+i(\omega_0-\omega)z} {E_0}\bar{E}_{1,y} + c.c.
\label{phi1b}
\end{equation}
Following this, we can write:
$$
a = a_0 e^{-i(\omega_0-\omega)t-i(\omega_0+\omega)z}
+a_0' e^{i(\omega_0+\omega)t+i(\omega_0-\omega)z}
 +c.c.
$$
with:
$$
a_0 = \frac{2\eta \bar{E_0}E_{1,y}}{4\omega_0 \omega +m_a^2}
\ \ \ \ \
a_0' = \frac{2\eta {E_0}\bar{E}_{1,y}}{-4\omega_0 \omega +m_a^2}
$$
Here, it is possible to inject this expression into equation (\ref{phi2})
and  derive the modified refractive index due to the presence of the ALP field $a$.
 We find the index along the $(x)$ axis is left unchanged while 
 $n_y$ is modified as:
\begin{equation}
n_y = 1+28 K  |E_0|^2+ \frac{ 4 \eta^2 m_a^2 |E_0|^2 }
{ m_a^4- (4\omega_0 \omega)^2}.
\label{nyalp}
\end{equation}
In the experimental conditions considered above, $\lambda=0.5$~~$\mu$m~$\sim 2.5$~eV, we obtain $(4\omega_0 \omega)^\frac{1}{2} \sim 5$ eV. Then, there is a resonant effect for ALPs of this mass, that will dominate the non-linear contribution in $K  |E_0|^2$.
For deep red of  $\lambda=0.7$~~$\mu$m~$\sim 1.75$~eV,
the resonance will be for $m_a=3.5$ eV.  This means 
that this experimental configuration is  interesting in order to obtain a  sensitivity (through a resonant effect) to ALP of masses of the order of eV, when there is the possibility to scan several values of laser wavelengths.
This strategy follows what has been developed for the LHC era: the search for ALPs as a resonant deviation in the light-by-light scattering.
In order to quantify this sensitivity, we need to compare the second term in equation (\ref{nyalp}), namely 
$\frac{ 4 \eta^2 m_a^2 |E_0|^2 }
{ m_a^4- (4\omega_0 \omega)^2}$
with 
the light-by-light term:
$28 K  |E_0|^2$. Thus, we need to compare:
 $\sqrt{K} \sim \frac{10^{-5} \mathrm{GeV}^{-1}}{\mathrm{eV}}$
 to 
 $\frac{\eta}{m_a} \cal{A}$, where $\cal{A}$ is the amplification factor due to the resonant effect: the better is the finesse of the cavities, the largest is $\cal{A}$. 
{%\color{red} 
 The existing and potential limits in the search for ALPs of masses of the eV or much below
 \cite{Asztalos:2011bm,Wagner:2010mi,Anastassopoulos:2017ftl,DellaValle:2015xxa,Ehret:2010mh} are of the order
 $\frac{\eta}{m_a} \sim \frac{10^{-10} \mathrm{GeV}^{-1}}{\mathrm{eV}}$ for ALPs masses of order of the eV. Therefore, we need an amplification factor $\cal{A}$ of order $10^5$ so that one may obtain competitive sensitivity using this technique, which is an experimental challenge. An amplification factor of $10^3$ to $10^4$  is already feasible with high quality cavities where the laser beams propagate.
Also, using astrophysical probes, some prospects have been extended for ALP in the eV range  down to photon-ALP coupling of $\frac{\eta}{m_a} \sim \frac{10^{-12} \mathrm{GeV}^{-1}}{\mathrm{eV}}$ \cite{Ringwald:2012hr}. However, the analysis using light-by-light scattering stays a complementary approach, that could already provide first results in a near future
\cite{Sarazin:2016zer}. 
 }

%cite{Ringwald:2012hr}

%%%%%%%%%%%%%%%
\section{Conclusions}\label{conclusions}
%%%%%%%%%%%%%%%

\noindent

We studied the discovery potential of ALPs coupled to photons by taking into account future opportunities in heavy-ion physics at the LHC, as well as in lower energies in the context of laser beam experiments. Lighter nuclei provide access to larger invariant masses for exclusive di-photon production in UPCs in comparison to heavier nuclei, with the disadvantage that the cross sections are not as large as with the standard lead nuclei used at the LHC. The interplay between these two effects were studied in this letter, which led to new projections on the ALP-photon coupling and mass plane, without additional assumptions on the ALP coupling to other SM particles. 

For our projections, we considered special runs with argon-argon collisions at the center-of-mass energy per nucleon pair of $7$ TeV with an integrated luminosity of $3$ pb$^{-1}$ and proton-lead collisions at the center-of-mass energy per nucleon pair of $8.16$ TeV with an integrated luminosity of $5$ pb$^{-1}$. We have presented in addition new projections considering a larger integrated luminosity of 10 nb$^{-1}$ of lead-lead collisions at a center-of-mass energy per nucleon pair of $5.02$ TeV to have a more complete comparison with the future heavy-ion physics program. 
We have found that argon-argon collisions have the potential to provide stringent constraints on the ALP-photon coupling for masses ranging between $100$~GeV and about $400$~GeV, which is precisely the region of parameter space that is difficult to access in proton-proton collisions and lead-lead collisions, and is thus complementary to both configurations. While proton-lead collisions have access to larger invariant masses than lead-lead collisions, the projections extracted from proton-lead collisions are severely limited by the expected amount of luminosity for this mode. The reach on the ALP coupling for larger invariant masses ends up being similar to what will be accessible in lead-lead collisions at $10$~nb$^{-1}$. 
At the early stages of this analysis, we considered other light ions at the LHC (oxygen, xenon). However, the amount of luminosity required in order to have competitive sensitivities with these ion species is forbiddingly large.

In addition, we have shown that ALPs of masses of the order of eV can be probed in laser beam experiments with already existing facilities. The existence of ALPs in this regime would manifest as a resonant-like effect in the refractive index of the vacuum. This strategy adds a new possibility to the set of experiments that are already performing searches of ALPs of eV masses or much below.

%%%%%%%%%%%%%%%
\section*{Acknowledgements}
%%%%%%%%%%%%%%%
 
\noindent
We thank   S.~Fichet, G.~Von Gersdorff, M. Saimpert for fruitful discussions during the writing of the manuscript. We also thank our colleagues who attended the LHC Forward Physics Working Group Meeting held at CERN in December 2018 for their valuable feedback and suggestions on our study.
CB thanks the financial support provided by the starting research grant of CR as a Distinguished Foundation Professor at the University of Kansas.

\providecommand{\href}[2]{#2}\begingroup\raggedright


\begin{thebibliography}{99}

%%%%%%%% electric dipole
\bibitem{Afach:2015sja}
  J.~M.~Pendlebury {\it et al.},
  %``Revised experimental upper limit on the electric dipole moment of the neutron,''
  Phys.\ Rev.\ D {\bf 92} (2015) no.9,  092003
  %doi:10.1103/PhysRevD.92.092003
  [arXiv:1509.04411 [hep-ex]].
  %%CITATION = doi:10.1103/PhysRevD.92.092003;%%
  %190 citations counted in INSPIRE as of 29 Jan 2019
%%%%%%%% 


%%%%%%%PQ
\bibitem{Peccei:1977ur}
  R.~D.~Peccei and H.~R.~Quinn,
  %``Constraints Imposed by CP Conservation in the Presence of Instantons,''
  Phys.\ Rev.\ D {\bf 16} (1977) 1791.
  %doi:10.1103/PhysRevD.16.1791
  %%CITATION = doi:10.1103/PhysRevD.16.1791;%%
  %2530 citations counted in INSPIRE as of 29 Jan 2019
\bibitem{Peccei:1977hh}
  R.~D.~Peccei and H.~R.~Quinn,
  %``CP Conservation in the Presence of Instantons,''
  Phys.\ Rev.\ Lett.\  {\bf 38} (1977) 1440.
  %doi:10.1103/PhysRevLett.38.1440
  %%CITATION = doi:10.1103/PhysRevLett.38.1440;%%
  %4617 citations counted in INSPIRE as of 29 Jan 2019
 \bibitem{Peccei:1977np}
  R.~D.~Peccei and H.~R.~Quinn,
  %``Some Aspects of Instantons,''
  Nuovo Cim.\ A {\bf 41} (1977) 309.
  %doi:10.1007/BF02730110
  %%CITATION = doi:10.1007/BF02730110;%%
  %39 citations counted in INSPIRE as of 29 Jan 2019

%%%%%%%%


%%%%%%%% CDM
\bibitem{Arias:2012az}
  P.~Arias, D.~Cadamuro, M.~Goodsell, J.~Jaeckel, %J.~Redondo and A.~Ringwald,
  JCAP {\bf 1206} (2012) 013
  [arXiv:1201.5902 [hep-ph]].
%%%%%%%% end CDM

%%%%%%%% 
%%%%%%%% superstring
%%%%%%%% 
\bibitem{Witten:1984dg}
  E.~Witten,
  %``Some Properties of O(32) Superstrings,''
  Phys.\ Lett.\  {\bf 149B} (1984) 351.
  %doi:10.1016/0370-2693(84)90422-2.
  %%CITATION = doi:10.1016/0370-2693(84)90422-2;%%
  %762 citations counted in INSPIRE as of 29 Jan 2019
\bibitem{Conlon:2006tq}
  J.~P.~Conlon,
  %``The QCD axion and moduli stabilisation,''
  JHEP {\bf 0605} (2006) 078
  %doi:10.1088/1126-6708/2006/05/078
  [hep-th/0602233].
  %%CITATION = doi:10.1088/1126-6708/2006/05/078;%%
  %188 citations counted in INSPIRE as of 25 Jan 2019
\bibitem{Svrcek:2006yi}
  P.~Svrcek and E.~Witten,
  %``Axions In String Theory,''
  JHEP {\bf 0606} (2006) 051
  %doi:10.1088/1126-6708/2006/06/051
  [hep-th/0605206].
  %%CITATION = doi:10.1088/1126-6708/2006/06/051;%%
  %621 citations counted in INSPIRE as of 25 Jan 2019
\bibitem{Arvanitaki:2009fg}
  A.~Arvanitaki, S.~Dimopoulos, S.~Dubovsky, N.~Kaloper and J.~March-Russell,
  %``String Axiverse,''
  Phys.\ Rev.\ D {\bf 81} (2010) 123530
  %doi:10.1103/PhysRevD.81.123530
  [arXiv:0905.4720 [hep-th]].
  %%CITATION = doi:10.1103/PhysRevD.81.123530;%%
  %562 citations counted in INSPIRE as of 25 Jan 2019
%%%%%%%% 
%%%%%%%% end superstring
%%%%%%%% 

%%%%%%%%  previous work: CR
\bibitem{us}
S. Fichet, G. von Gersdorff, O. Kepka, B. Lenzi, C. Royon, M. Saimpert, Phys. Rev. D {\bf 89} (2014) 114004;
S. Fichet, G. von Gersdorff, B. Lenzi, C. Royon, M. Saimpert, JHEP {\bf 1502} (2015) 165;
S. Fichet, G. von Gersdorff, C. Royon, Phys.Rev. D {\bf 93} (2016) no.7, 075031;
S. Fichet, G. von Gersdorff, C. Royon, Phys. Rev. Lett. {\bf 116} (2016) no.23, 231801.
%%%%%%%% 
  
%%%%%%%%%%intro to LbL LHC
\bibitem{dEnterria:2013zqi}
  D.~d'Enterria and G.~G.~da Silveira,
  %``Observing light-by-light scattering at the Large Hadron Collider,''
  Phys.\ Rev.\ Lett.\  {\bf 111} (2013) 080405
   Erratum: [Phys.\ Rev.\ Lett.\  {\bf 116} (2016) no.12,  129901]
  %doi:10.1103/PhysRevLett.111.080405, 10.1103/PhysRevLett.116.129901
  [arXiv:1305.7142 [hep-ph]].
  %%CITATION = doi:10.1103/PhysRevLett.111.080405, 10.1103/PhysRevLett.116.129901;%%
  %91 citations counted in INSPIRE as of 31 Jan 2019
    
%%%%%%%%%%ATLAS
\bibitem{Aaboud:2017bwk}
  M.~Aaboud {\it et al.} [ATLAS Collaboration],
  %``Evidence for light-by-light scattering in heavy-ion collisions with the ATLAS detector at the LHC,''
  Nature Phys.\  {\bf 13} (2017) no.9,  852
  %doi:10.1038/nphys4208
  [arXiv:1702.01625 [hep-ex]].
  %%CITATION = doi:10.1038/nphys4208;%%
  %72 citations counted in INSPIRE as of 31 Jan 2019
%%%%CMS
\bibitem{Sirunyan:2018fhl}
  A.~M.~Sirunyan {\it et al.} [CMS Collaboration],
  %``Evidence for light-by-light scattering and searches for axion-like particles in ultraperipheral PbPb collisions at $\sqrt{s_\mathrm{NN}} =$ 5.02 TeV,''
  arXiv:1810.04602 [hep-ex].
  %%CITATION = ARXIV:1810.04602;%%
  %8 citations counted in INSPIRE as of 29 Jan 2019

    
%%%%knapen
\bibitem{Knapen:2016moh}
  S.~Knapen, T.~Lin, H.~K.~Lou and T.~Melia,
  %``Searching for Axionlike Particles with Ultraperipheral Heavy-Ion Collisions,''
  Phys.\ Rev.\ Lett.\  {\bf 118} (2017) no.17,  171801
  %doi:10.1103/PhysRevLett.118.171801
  [arXiv:1607.06083 [hep-ph]],
  %``LHC limits on axion-like particles from heavy-ion collisions,''
  arXiv:1709.07110 [hep-ph].
  %%CITATION = ARXIV:1709.07110;%%
  %10 citations counted in INSPIRE as of 29 Jan 2019
  %%CITATION = doi:10.1103/PhysRevLett.118.171801;%%
  %40 citations counted in INSPIRE as of 29 Jan 2019

%%%%%%cristian
\bibitem{Baldenegro:2018hng}
C.~Baldenegro, S.~Fichet, G.~von Gersdorff and C.~Royon,
  %``Searching for axion-like particles with proton tagging at the LHC,''
  JHEP {\bf 1806} (2018) 131;
  %doi:10.1007/JHEP06(2018)131
  %[arXiv:1803.10835 [hep-ph]].
  %%CITATION = doi:10.1007/JHEP06(2018)131;%%
  %12 citations counted in INSPIRE as of 13 Mar 2019
JHEP {\bf 1706} (2017) 142;
E. Chapon, C. Royon, O. Kepka, Phys.Rev. D{\bf 81} (2010) 074003;
O. Kepka, C. Royon, Phys.Rev. D{\bf 78} (2008) 073005.
 
%%%%%LbL orsay
\bibitem{Sarazin:2016zer}
  X.~Sarazin, F.~Couchot, A.~Djannati-Ataï, O.~Guilbaud, S.~Kazamias, M.~Pittman and M.~Urban,
  %``Refraction of light-by-light in vacuum,''
  Eur.\ Phys.\ J.\ D {\bf 70} (2016) no.1,  13
  %doi:10.1140/epjd/e2015-60428-5
  [arXiv:1507.07959 [physics.optics]].
  %%CITATION = doi:10.1140/epjd/e2015-60428-5;%%
  %3 citations counted in INSPIRE as of 21 Feb 2019

%%%King
\bibitem{King:2012aw}
  B.~King and C.~H.~Keitel,
  %``Photon-photon scattering in collisions of laser pulses,''
  New J.\ Phys.\  {\bf 14} (2012) 103002
  %doi:10.1088/1367-2630/14/10/103002
  [arXiv:1202.3339 [hep-ph]].
  %%CITATION = doi:10.1088/1367-2630/14/10/103002;%%
  %34 citations counted in INSPIRE as of 22 Feb 2019

%%%%the nice paper...
\bibitem{Takahashi:2018uut}
  T.~Takahashi {\it et al.},
  %``Light-by-light scattering in a photon–photon collider,''
  Eur.\ Phys.\ J.\ C {\bf 78} (2018) no.11,  893
  %doi:10.1140/epjc/s10052-018-6364-1
  [arXiv:1807.00101 [hep-ex]].
  %%CITATION = doi:10.1140/epjc/s10052-018-6364-1;%%
\bibitem{Bogorad:2019pbu}
  Z.~Bogorad, A.~Hook, Y.~Kahn and Y.~Soreq,
  %``Probing ALPs and the Axiverse with Superconducting Radiofrequency Cavities,''
  arXiv:1902.01418 [hep-ph].
  %%CITATION = ARXIV:1902.01418;%%
  %2 citations counted in INSPIRE as of 12 Mar 2019


%%%%GDR  
\bibitem{Auerbach:1983hld}
  N.~Auerbach and A.~Klein,
  %``A microscopic theory of giant electric isovector resonances,''
  Nucl.\ Phys.\ A {\bf 395} (1983) 77.
  %%doi:10.1016/0375-9474(83)90090-8
  %%CITATION = doi:10.1016/0375-9474(83)90090-8;%%
  %91 citations counted in INSPIRE as of 04 Mar 2019
  
  
%%%%%%%%%%YR
\bibitem{Citron:2018lsq}
  Z.~Citron {\it et al.},
  %``Future physics opportunities for high-density QCD at the LHC with heavy-ion and proton beams,''
  arXiv:1812.06772 [hep-ph].
  %%CITATION = ARXIV:1812.06772;%%
  %7 citations counted in INSPIRE as of 29 Jan 2019
  
%%%%%%FPMC  
\bibitem{Boonekamp:2011ky}
  M.~Boonekamp, A.~Dechambre, V.~Juranek, O.~Kepka, M.~Rangel, C.~Royon and R.~Staszewski,
  %``FPMC: A Generator for forward physics,''
  arXiv:1102.2531 [hep-ph].
  %%CITATION = ARXIV:1102.2531;%%
  %57 citations counted in INSPIRE as of 30 Jan 2019


%%%%%%%%%starlight
\bibitem{Klein:2016yzr}
  S.~R.~Klein, J.~Nystrand, J.~Seger, Y.~Gorbunov and J.~Butterworth,
  %``STARlight: A Monte Carlo simulation program for ultra-peripheral collisions of relativistic ions,''
  Comput.\ Phys.\ Commun.\  {\bf 212} (2017) 258
  %doi:10.1016/j.cpc.2016.10.016
  [arXiv:1607.03838 [hep-ph]].
  %%CITATION = doi:10.1016/j.cpc.2016.10.016;%%
  %66 citations counted in INSPIRE as of 30 Jan 2019

\bibitem{Klusek-Gawenda:2016euz}
  M.~Kłusek-Gawenda, P.~Lebiedowicz and A.~Szczurek,
  %``Light-by-light scattering in ultraperipheral Pb-Pb collisions at energies available at the CERN Large Hadron Collider,''
  Phys.\ Rev.\ C {\bf 93} (2016) no.4,  044907
  %doi:10.1103/PhysRevC.93.044907
  [arXiv:1601.07001 [nucl-th]].
  %%CITATION = doi:10.1103/PhysRevC.93.044907;%%
  %39 citations counted in INSPIRE as of 21 Feb 2019


%%%%%%%detector simulation  
\bibitem{Aad:2009wy}
  G.~Aad {\it et al.} [ATLAS Collaboration],
  %``Expected Performance of the ATLAS Experiment - Detector, Trigger and Physics,''
  arXiv:0901.0512 [hep-ex].
  %%CITATION = ARXIV:0901.0512;%%
  %2101 citations counted in INSPIRE as of 30 Jan 2019



%%%superchic
\bibitem{Harland-Lang:2018iur}
  L.~A.~Harland-Lang, V.~A.~Khoze and M.~G.~Ryskin,
  %``Exclusive LHC physics with heavy ions: SuperChic 3,''
  Eur.\ Phys.\ J.\ C {\bf 79} (2019) no.1,  39
  %doi:10.1140/epjc/s10052-018-6530-5
  [arXiv:1810.06567 [hep-ph]].
  %%CITATION = doi:10.1140/epjc/s10052-018-6530-5;%%
  %7 citations counted in INSPIRE as of 22 Feb 2019


%%%%%%Bauer
\bibitem{Bauer:2017ris}
M.~Bauer, M.~Neubert and A.~Thamm, {\emph{JHEP}
  {\bfseries 12} (2017) 044}
  [\href{https://arxiv.org/abs/1708.00443}{{\ttfamily 1708.00443}}].

\bibitem{Jaeckel:2015jla}
  J.~Jaeckel and M.~Spannowsky,
  %``Probing MeV to 90 GeV axion-like particles with LEP and LHC,''
  Phys.\ Lett.\ B {\bf 753} (2016) 482
  %%doi:10.1016/j.physletb.2015.12.037
  [arXiv:1509.00476 [hep-ph]];
  %%CITATION = doi:10.1016/j.physletb.2015.12.037;%%
  %69 citations counted in INSPIRE as of 12 Mar 2019
  J.~Jaeckel, M.~Jankowiak and M.~Spannowsky,
  %``LHC probes the hidden sector,''
  Phys.\ Dark Univ.\  {\bf 2} (2013) 111
  %%doi:10.1016/j.dark.2013.06.001
  [arXiv:1212.3620 [hep-ph]].
  %%CITATION = doi:10.1016/j.dark.2013.06.001;%%
  %91 citations counted in INSPIRE as of 12 Mar 2019



%%%SLAC experiment beam dump  
\bibitem{PhysRevLett.59.755}
E.~M. Riordan, M.~W. Krasny, K.~Lang, P.~de~Barbaro, A.~Bodek, S.~Dasu et~al.,
  {\emph{Phys. Rev. Lett.}
  {\bfseries 59} (1987) 755}.

\bibitem{PhysRevD.38.3375}
J.~D. Bjorken, S.~Ecklund, W.~R. Nelson, A.~Abashian, C.~Church, B.~Lu et~al.,
  {\emph{Phys.
  Rev. D} {\bfseries 38} (1988) 3375}.

\bibitem{Döbrich2016}
B.~D{\"o}brich, J.~Jaeckel, F.~Kahlhoefer, A.~Ringwald and K.~Schmidt-Hoberg,
  
  {\emph{Journal of High Energy Physics} {\bfseries 2016} (2016) 18}.


%%%%babar belle
\bibitem{PhysRevD.51.2053}
R.~Balest et~al.,
 {\emph{Phys. Rev.
  D} {\bfseries 51} (1995) 2053}.

\bibitem{delAmoSanchez:2010ac}
{\scshape BaBar} collaboration, P.~del Amo~Sanchez et~al., {\emph{Phys. Rev. Lett.}
  {\bfseries 107} (2011) 021804}
  [\href{https://arxiv.org/abs/1007.4646}{{\ttfamily 1007.4646}}].

%%%%%CLS
\bibitem{Read:2002hq}
  A.~L.~Read,
  %``Presentation of search results: The CL(s) technique,''
  J.\ Phys.\ G {\bf 28} (2002) 2693.
  %doi:10.1088/0954-3899/28/10/313
  %%CITATION = doi:10.1088/0954-3899/28/10/313;%%
  %2397 citations counted in INSPIRE as of 28 Feb 2019

{%\color{red}
%%% tri-photons
\bibitem{Mimasu:2014nea}
  K.~Mimasu and V.~Sanz,
  %``ALPs at Colliders,''
  JHEP {\bf 1506} (2015) 173
  %%doi:10.1007/JHEP06(2015)173
  [arXiv:1409.4792 [hep-ph]].
  %%CITATION = doi:10.1007/JHEP06(2015)173;%%
  %48 citations counted in INSPIRE as of 11 May 2019
}

\bibitem{scott} Robertson S.J., {\it private communication}.

%%%%%ADMX
\bibitem{Asztalos:2011bm}
  S.~J.~Asztalos {\it et al.} [ADMX Collaboration],
  %``Design and performance of the ADMX SQUID-based microwave receiver,''
  Nucl.\ Instrum.\ Meth.\ A {\bf 656} (2011) 39
  %doi:10.1016/j.nima.2011.07.019
  [arXiv:1105.4203 [physics.ins-det]].
  %%CITATION = doi:10.1016/j.nima.2011.07.019;%%
  %50 citations counted in INSPIRE as of 21 Feb 2019
  
\bibitem{Wagner:2010mi}
  A.~Wagner {\it et al.} [ADMX Collaboration],
  %``A Search for Hidden Sector Photons with ADMX,''
  Phys.\ Rev.\ Lett.\  {\bf 105} (2010) 171801
  %doi:10.1103/PhysRevLett.105.171801
  [arXiv:1007.3766 [hep-ex]].
  %%CITATION = doi:10.1103/PhysRevLett.105.171801;%%
  %65 citations counted in INSPIRE as of 21 Feb 2019

%%%%CAST
\bibitem{Anastassopoulos:2017ftl}
  V.~Anastassopoulos {\it et al.} [CAST Collaboration],
  %``New CAST Limit on the Axion-Photon Interaction,''
  Nature Phys.\  {\bf 13} (2017) 584
  %doi:10.1038/nphys4109
  [arXiv:1705.02290 [hep-ex]].
  %%CITATION = doi:10.1038/nphys4109;%%
  %109 citations counted in INSPIRE as of 21 Feb 2019

%%%PVLAS
\bibitem{DellaValle:2015xxa}
  F.~Della Valle, A.~Ejlli, U.~Gastaldi, G.~Messineo, E.~Milotti, R.~Pengo, G.~Ruoso and G.~Zavattini,
  %``The PVLAS experiment: measuring vacuum magnetic birefringence and dichroism with a birefringent Fabry–Perot cavity,''
  Eur.\ Phys.\ J.\ C {\bf 76} (2016) no.1,  24
  %doi:10.1140/epjc/s10052-015-3869-8
  [arXiv:1510.08052 [physics.optics]].
  %%CITATION = doi:10.1140/epjc/s10052-015-3869-8;%%
  %68 citations counted in INSPIRE as of 21 Feb 2019
B
%%%%%%ALPS
\bibitem{Ehret:2010mh}
  K.~Ehret {\it et al.},
  %``New ALPS Results on Hidden-Sector Lightweights,''
  Phys.\ Lett.\ B {\bf 689} (2010) 149
  %doi:10.1016/j.physletb.2010.04.066
  [arXiv:1004.1313 [hep-ex]].
  %%CITATION = doi:10.1016/j.physletb.2010.04.066;%%
  %221 citations counted in INSPIRE as of 21 Feb 2019

{%\color{red}
%%%%%%ALPs cosmology (star cooling)
\bibitem{Ringwald:2012hr}
  A.~Ringwald,
  %``Exploring the Role of Axions and Other WISPs in the Dark Universe,''
  Phys.\ Dark Univ.\  {\bf 1} (2012) 116
  %doi:10.1016/j.dark.2012.10.008
  [arXiv:1210.5081 [hep-ph]].
  %%CITATION = doi:10.1016/j.dark.2012.10.008;%%
  %170 citations counted in INSPIRE as of 11 May 2019
}


\end{thebibliography}
\end{document}